\documentclass[12pt]{article}
\pdfoutput=1
 \usepackage{latexsym, graphicx} 

\usepackage[colorlinks=true, linkcolor=blue, bookmarks=true]{hyperref}

\makeatletter
\renewcommand\section{\@startsection {section}{1}{\z@}%
                                   {-3.5ex \@plus -1ex \@minus -.2ex}
                                   {2.3ex \@plus.2ex}%
                                   {\normalfont\large\bfseries}}
\renewcommand\subsection{\@startsection{subsection}{2}{\z@}%
                                     {-3.25ex\@plus -1ex \@minus -.2ex}%
                                     {1.5ex \@plus .2ex}%
                                     {\normalfont\bfseries}}
\makeatother
\newcommand{\beq}{\begin{equation}}
\newcommand{\eeq}{\end{equation}}
\newcommand{\ber}{\begin{array}}
\newcommand{\eer}{\end{array}}
\newcommand{\D}{{\cal D}}

\newcommand{\ssty}{\scriptstyle}
\newcommand{\dsty}{\displaystyle}

\newcommand{\ena}{\end{eqnarray}}
\newcommand{\beqa}{\begin{eqnarray}}
\newcommand{\eeqa}{\end{eqnarray}}
\newcommand{\bea}{\begin{eqnarray}}
\newcommand{\eea}{\end{eqnarray}}

\newcommand{\be}{\begin{equation}}
\newcommand{\ee}{\end{equation}}

\begin{document}
\begin{titlepage}
\begin{flushright}
\phantom{arXiv:yymm.nnnn}
\end{flushright}
\vfill
\begin{center}
{\Large\bf Microscopic approach to string gas cosmology}    \\
\vskip 15mm
{\large Oleg Evnin}
\vskip 7mm
{\em Department of Physics, Faculty of Science, Chulalongkorn University,\\
Thanon Phayathai, Pathumwan, Bangkok 10330, Thailand}

\vskip 3mm
{\small\noindent  {\tt oleg.evnin@gmail.com}}
\vskip 10mm
\end{center}
\vfill

\begin{center}
{\bf ABSTRACT}\vspace{3mm}
\end{center}
In this contribution to the proceedings of  the Conference on Modern Physics of Compact Stars and Relativistic Gravity in Yerevan, Armenia (September 18-21, 2013), I review recent work attempting to give a fundamental definition to string evolution in a dynamical, fully compact universe, and present a sketch of how the resulting formalism can be used for addressing questions of phenomenological significance in the field of string gas cosmology.

\vfill

\end{titlepage}

\section{Introduction}

Considerations of the dynamics of string gases in fully compact cosmological spaces have produced a number of attractive phenomenological conjectures, starting with the pioneering work \cite{bv} (for reviews, see \cite{rv1,rv2}). Besides providing a possible alternative to the inflationary pardigm in relation to the horizon problem, etc, the set-up is quite unique in its potential ability to explain the number of macroscopic spatial dimensions of the universe, which equals three observationally. As emphasized in \cite{bv}, the special kinematics of stretched string scattering is three dimensions is likely to enforce expansion of exactly three macroscopic dimensions, no matter what additional compact spatial dimensions are present in the universe.
(Note that `strings' in our present context mean the fundamental quantum relativistic strings of string theory, rather than effective solitonic strings in field theories, for example. Though similar effects may possibly be induced by other stringy states, the work reviewed here focuses on string gas cosmology scenarios based on fundamental strings.)

Since the introduction of the key ideas in \cite{bv}, much work has been done in an attempt to implement this type of string cosmology scenarios at a more quantitative level, with mixed results. For a sampling of publications, see \cite{sg2,sg3,sg4,sg5}, with more references to earlier work available in the reviews \cite{rv1,rv2}. The most recent publications claim a successful implementation of the scenario, with only three macroscopic spatial dimensions expanding.

A common feature of most of the above considerations is that they rely on a `macroscopic' thermodynamic description of the string gas (followed by numerical simulations of the cosmological expansion it induces). As one is talking about universes whose size is of order of the string scale and whose energy density is of order of the string scale (corresponding to the Hagedorn temperature of string theory), one is typically dealing with just a few string quanta moving in a string-sized universe. One is then, at best, in a `mesoscopic' regime, where the validity of thermodynamical treatment would require extra justification. Coupled with the inherent complexity of the string dynamics and the difficulty of giving its precise thermodynamical description, this may give some attraction to a fully `microscopic' alternative: working with the string gas in terms of pure quantum states. 

Even apart from the practical advantages or disadvantages of such a fully microscopic approach to string gases in small compact universes, it is important to ascertain that there is a way to implement this type of fundamental description in lines with the general principles of string theory. As it turns out, na\"\i ve applications of the usual perturbative string theory in fully compact spaces fail for a number of reasons. The most prominent of these failures are the infrared divergences arising when all spatial dimensions have been compactified (and related to the special kinematics in 0+1 dimensions). These divergences are intimately related \cite{procpad} to those emerging in the soliton recoil problem in field theory \cite{raj,lee} and the D0-brane recoil problem in string theory \cite{pt,hk,d0}. The infrared divergences are a reflection of the gravitational backreaction induced by even one string moving in a compact cosmological space (in a non-compact space, this backreaction does not affect the asymptotic regions, hence the background remains close to the unperturbed one almost everywhere). Due to the large ratio of the Planck and string scales, the divergences only affect the string perturbative expansion starting from the first subleading order (again, this is similar to the recoil problems). One needs to obtain a reformulation of the usual perturbative string theory free from the infrared problems, if one is to have a fundamental microscopic description of strings in a fully compact universe.

The problem of reorganizing the usual string perturbation theory in compact spaces has been considered in \cite{compact}. It has been suggested that a functional integral over the modes of the background receiving large backreaction should be introduced in addition to the usual functional integral over string worldsheets. Such a treatment deviates from the usual paradigm of formulating string theory in a fixed classical background, with massless string quanta representing small deformations of this background. One reason for this unusual treatment is that, in the presence of strong backreaction, deviations from the unperturbed background are not small. This type of modified string perturbation theory is modelled on the solution to the D0-brane recoil problem in \cite{d0}, which is in turn modelled on the Christ-Lee perturbation theory for the field-theoretical soliton recoil problem \cite{lee}. Note that, technically, these discussions are phrased for bosonic strings moving in a toroidal universe, though the core issues are completely general and in no way dependent on the presence or absence of supersymmetry, or the detailed topology, but only on compactness of the spatial manifold.

In this contribution, I shall review the formalism of \cite{compact}, in a way somewhat different from how it was presented in the original publication. I will then give a brief sketch of some practical computations this formalism can be used for.

\section{The formalism}

I shall start by simply presenting the proposed expression for the evolution amplitude, subsequently providing justifications and clarifications. For a more systematic and detailed approach, one can consult \cite{compact}.

What amplitudes could one define in our cosmological setting? Since the space is fully compact, the usual notion of S-matrix clearly does not apply. One should therefore resort to some sort of finite-time transition amplitudes. The question of finite-time transition amplitudes in string theory has not been settled even in non-compact Minkowski space, but a formal definition has been given, for example, in \cite{cmp}. One can try to extend this definition to the case when the background is dynamical rather than fixed. Due to the time-reparametrization invariance of the background space-time, it will only make sense to talk about the amplitudes for the universe to have two cross-sections with specified spatial properties and string contents (without any specification of the `moments of time' when such configurations occur, since that notion is not reparametrization-invariant). This is very similar to the kind of amplitudes considered in minisuperspace models of quantum cosmology \cite{hh,halliwell, isham, marolf}.

\begin{figure}
\begin{center}
\includegraphics[width=5cm, bb=200 580 360 720]{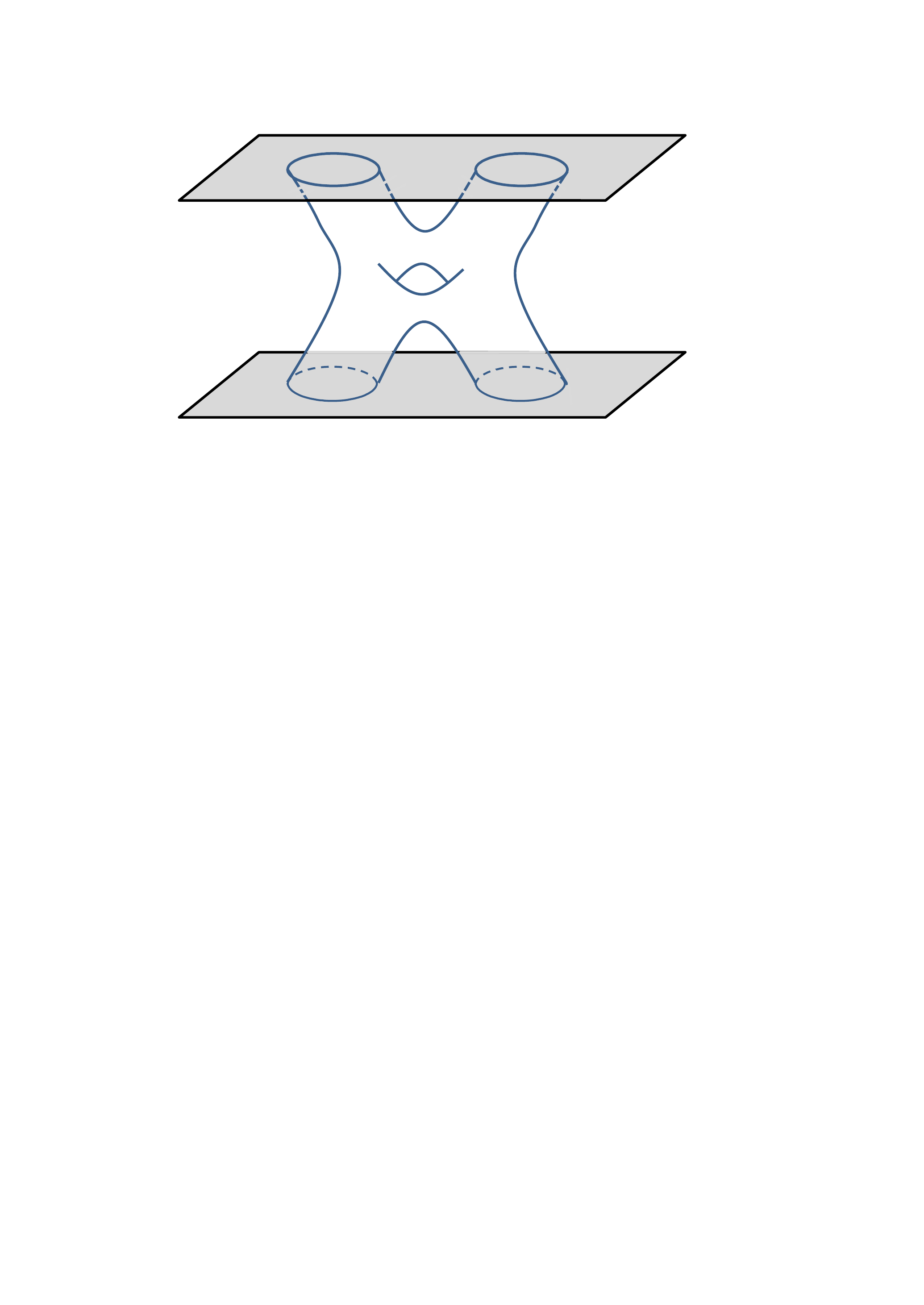}
\end{center}
\caption{A string worldsheet cut by the two observation moments in a dynamical universe.}
\label{wscut}
\end{figure}

We shall integrate over the following class of backgrounds for the metric, the 2-form and the dilaton (the justification comes from analyzing which massless field modes contribute to the infrared divergences in a na\"\i ve formulation of perturbative string theory in a fixed space-time background):
\beq
\ber{l}
\dsty ds^2=-G_{00}(t)dt^2+2G_{0i}(t)dtdx^i+G_{ij}(t)dx^idx^j,\qquad x^i\in [0,2\pi), \vspace{2mm}\\
\dsty B_{\mu\nu}=B_{\mu\nu}(t),\qquad \Phi=\Phi(t).
\eer
\label{IRmodes}
\eeq
In the language of the minisuperspace literature, this is a compactified Bianchi I universe coupled to the dilaton and the 2-form.
$G_{00}$ and $G_{0i}$ can be gauged away completely, except for a single gauge-invariant number, $\int dt \sqrt{G_{00}}$. We shall suppress $B_{\mu\nu}$ and $\Phi$ in the formulas and discussions below for compactness, but it is understood that they are also included, whenever the metric $G_{ij}$ appears. One finally specifies the values of the background fields (\ref{IRmodes}), which we denote $G^{(1)}$ and $G^{(2)}$, on the two observation hypersurfaces, together with the (loop-like) configurations of closed strings, which we denote $\ell^{(1)}$ and $\ell^{(2)}$, on the same observation hypersurfaces. (A pictorial representation of the slice of the worldsheet between the two observation hypersurfaces is given in Fig.~\ref{wscut}.)
One then defines the following transition amplitude.
\beq
\langle G^{(2)},\ell^{(2)}|G^{(1)},\ell^{(1)}\rangle=\int\limits_{-\infty}^{\infty}dT\hspace{-6mm}\int\limits_{\ber{l}\ssty G_{ij}(0)=G^{(1)}_{ij}\vspace{-0.5mm}\\ \ssty G_{ij}(T)=G^{(2)}_{ij}\eer} \hspace{-8mm}\D G_{ij}(t)\, e^{iS_{\rm ms}}\left(\prod\limits_{\ell^{(1)}_m\ell^{(2)}_n}\int_{Diff(S^1)}\right)\int\limits_{\ell^{(1)}_m\ell^{(2)}_n} \D X e^{-S_X},
\label{masterclosed}
\eeq
Here, $S_{\rm ms}$ is the low-energy string theory action with the background (\ref{IRmodes}) substituted in it, and $S_X$ is the (non-linear sigma-model) action for a string worldsheet propagating in (\ref{IRmodes}). Both actions can be found in chapter 3 of \cite{polchvol1}, for instance.

Note that (\ref{masterclosed}) has a rather peculiar structure, compared to ordinary quantum-mechanical transition amplitudes (``integral of the exponential of the action constrained by the boundary conditions''). Namely, one has to integrate over $T$, the end-time of the experiment, and over all reparametrizations of the boundary loops of the string worldsheet ($\ell^{(1)}$ and $\ell^{(2)}$), which is represented by $\int_{Diff(S^1)}$ in ($\ref{masterclosed}$). The first of the two integrals is mandated by the time-reparametrization invariance of the background, and indeed it eliminates the unphysical dependence on the observation time. Such amplitude structures have been thoroughly considered in the minisuperspace literature, see \cite{halliwell,marolf}. The integral over boundary reparametrizations is mandated by the diff-invariance on the string worldsheet, and eliminates the unphysical dependence on how the boundary loops of the string worldsheet are parametrized. Such integrals have appeared in considerations of off-shell string amplitudes in Minkowski space, see \cite{cmp}.

Having formally defined the amplitude in a closed form we shall now review the problems in conventional perturbative string theory applied to compact spaces that lead one to consider such amplitudes, as well as various consistency checks that the amplitude (\ref{masterclosed}) satisfies.

\subsection{Motivation}

The functional integration over a class of backgrounds in the basic amplitude (\ref{masterclosed}) may appear as a rather odd departure from the usual perturbative string theory (which only includes a functional integral over worldsheets), and it is important to explain why the usual formulation of string theory fails for compact spaces.

The most conservative approach to string theory in compact spaces would be to consider a toroidal space with a non-compact time and a flat Minkowski metric, and study how strings propagate in this background (this is simply a full spatial compactification of the usual string theory in Minkowski space). Naturally, amplitudes will be given by a sum over worldsheet topologies, with higher topologies (worldsheets with more handles) weighted by higher powers of the string coupling. The first sub-leading correction will be given by worldsheets of toroidal topology, and by the general rules of string theory, these amplitudes must be integrated over the modular parameter of the torus. There is a limit in this modular parameter integration that corresponds to the torus becoming thin and long. In this limit, an infrared divergence will arise in a fully compact target space, invalidating string perturbation theory.

We shall not derive the divergence accurately here, and only give a qualitative hint on how it arises. (The reader may consult \cite{compact} for further details.) Basically, the limit of the long thin torus worldsheet can be expressed through sphere diagrams augmented by an additional loop with different string states running in the loop. In particular, there will be contributions from the zeroth Kaluza-Klein (uniform) modes of massless string fields on the spatial torus running in the loop. The contribution of the infrared region of the loop momentum to these amplitudes will simply be proportional to
\beq
\int\limits_{k\approx 0} \frac{dk}{k^2},
\eeq
where $1/k^2$ is nothing but the massless particle propagator. This expression will be divergent on account of the propagator singularity at $k=0$. (If more than one spatial dimension were decompactified, the above integral would be convergent around $k=0$, because of the measure $dk$. The case of exactly one spatial dimension is subtle, see \cite{local}. There are no divergences from large values of $k$ in any situation in string theory, because of the known UV-finiteness.)

The divergence we have displayed is very robust, and essentially depends on kinematics in $0+1$ dimensions. Identical divergences arise in the D0-brane recoil problem \cite{pt,d0}.

\subsection{Justification}

Having observed the divergence trouble in the usual string perturbation theory put on a compact space, one might look for possible cures. An important source of inspiration comes from the Christ-Lee perturbation theory for recoiling field-theoretical solitons \cite{lee}, which cures divergences of similar algebraic structure and physical origin by reorganizing the perturbative expansion of field theory in such a way that a functional integral over the center-of-mass position of the soliton appears explicitly. Since in field theory the perturbative expansion is derived from the underlying Lagrangian formulation, the justification of the Christ-Lee perturbation theory is exact.

In string theory, one does not have the advantage of being able to derive the perturbative expansion from a more fundamental underlying framework. Nonetheless, one might try to imitate the Christ-Lee perturbation theory in the language of perturbative string theory and see whether it works. For the problem of D0-brane recoil, this was done in \cite{d0}, and it was shown that the divergences cancel as a result. The amplitude (\ref{masterclosed}) is an attempt to extend the same kind of treatment to the case of strings in a compact target space. (The similarities of the infrared divergences one has to deal with in all these cases have already been pointed out.)

One might find it conterintuitive that an integral over (a class of) background deformations is included in the amplitude (\ref{masterclosed}), whereas it is usually thought that such modes are already described by excitations of strings. The conventional intuition is, however, unlikely to apply to the particular modes we have included. For example, the usual string-theoretic description of the initial and final states of the zeroth Kaluza-Klein modes of the massless fields would be in terms of massless string quanta in different on-shell states. However, if all spatial dimensions are compactified, zeroth Kaluza-Klein modes have zero momenta in all spatial directions, and hence on-shell their momentum is exactly zero. It is difficult to imagine how one would describe initial and final states of the minisuperspace cosmology (\ref{IRmodes}) in terms of different numbers of massless string quanta at exactly zero momentum. Clearly, something is wrong with the usual string theory conventions when the number of non-compact dimensions becomes small!

\subsection{Finiteness properties}

The ultimate check for whether the proposed amplitudes (\ref{masterclosed}) make sense is in the way they resolve the infrared issues of the usual perturbative string theory on a compact space.

For the case of D0-brane recoil, which has been used here as a prototype for our treatment of the compact space case, the resolution is quite straightforward \cite{d0}. The modification of the formalism we introduce (functional integration over a set of background modes) produces a new perturbatively small divergence on the lowest genus worldsheet that cancels the divergence in the modular integration on the next-to-lowest genus worldsheet inherited from the standard string theory. Such a cancellation pattern is known as the Fischler-Susskind mechanism.

The situation for the present amplitudes (\ref{masterclosed}) is somewhat more subtle, in the sense that they are essentially finite-time transition amplitudes (there is no obvious way to talk about infinite time scattering in a compact space). For finite time transition amplitudes, one cannot expect infrared divergences in the first place (the space is compact and the time is cut off to a finite duration). Hence, amplitudes (\ref{masterclosed}) are automatically finite in this straightforward na\"\i ve sense.

If one looks a little deeper, one discovers that there are extremely abrupt dependences on the worldsheet and modular integration cut-offs in different contributions to (\ref{masterclosed}), inherited from the corresponding infrared divergences on an infinite time interval. These abrupt dependences would require cut-offs to be sent to numbers of the sort of $10^{-100}$ for perfectly reasonable amplitudes before a cut-off-independent limit is reached (in each of the terms of the amplitude separately). However, the dangerous cut-off-dependent terms cancel between the different contributions to the amplitude exactly in the manner of the Fischler-Susskind mechanism described above.

Overall, one discovers that (\ref{masterclosed}) addresses the infrared problems of the usual formulation of string theory on a compact space. Technical details can be recovered from \cite{compact}.

\section{The observables}

Armed with the basic definition of transition amplitudes (\ref{masterclosed}), it is natural to ask how they can be used to address questions of phenomenological significance in string gas cosmology, either analytically or numerically.

Even though much of the work in \cite{compact} analyzing the validity of the newly constructed formalism was at first sub-leading order in the string coupling (torus topology worldsheets), and hence technically rather involved, one should expect that the most important practical qualitative predictions will emerge at leading order (sphere topology worldsheets) and will be computationally much more straightforward in a number of ways. The general idea is to specify a certain set of strings on the two observation hypersurfaces (Fig. \ref{wscut}) and see how their presence affects the cosmological expansion, i.e., how the amplitudes for cosmological evolution without any strings get modified when strings are present.

Further simplifications arise from the fact that, for transitions between universes whose size is comparable to the string scale, which is the regime relevant in string cosmology, the functional integral over the background modes in (\ref{masterclosed}) is in a semi-classical regime. (This is a reflection of the large ratio between the Planck and string scales in perturbative string theory.) The leading contribution to the amplitude (\ref{masterclosed}) will then arise from strings propagating in a fixed classical background connecting the initial and final background configurations. Furthermore, it is likely that an adequate estimate for the worldsheet functional integral will be given simply be the exponential of the minimal possible area of the worldsheet connecting the initial and final configurations of the strings. The background solutions for Bianchi I gravity-dilaton system are known as Kasner-dilaton solutions, and with the above picture in mind, one should expect the leading string backreaction effects to be expressible through fairly straightforward geometrical quantities in the Kasner-dilaton spacetimes.

It remains to be seen up to which point the computations sketched here can be carried out in practice, and what light they shed, for example, on the question of the number of macroscopic dimensions expanding. (I.e., one may start with a string-sized universe configuration, in which the expansion of all dimensions of the torus is blocked by winding strings, and see how likely it is to discover another cross-section of the same universe, in which a certain number of winding strings have annihilated and a certain number of dimensions have expanded to much bigger sizes.)

\section{Acknowledgments}
I would like to thank the organizers of the Conference on Modern Physics of Compact Stars and Relativistic Gravity in Yerevan, Armenia (September 18-21, 2013), and specifically Armen Sedrakian. I would also like to thank Jarah Evslin, Chethan Krishnan, Karapet Mkrtchyan and Levon Pogosian for stimulating communication during the conference, and Ben Craps and Anatoly Konechny for collaboration on the research reviewed here. This work has been supported in part by Ratchadaphisek Sompote Endowment Fund.

\end{document}